\documentclass[useAMS,usenatbib]{mn2e}
\usepackage{graphicx}
\usepackage{epsfig}
\usepackage{multirow} 
\usepackage{subfigure}
\def\lesssim{\la}
\def\gtrsim{\ga}

\title[]{Stellar Migration by Short Lived Density Peaks Arising from Interference of Spiral Density 
Waves in an N-body Simulation}
\author[]{Justin Comparetta,  \& Alice C. Quillen     \\
{Department of Physics and Astronomy, University of Rochester, Rochester, NY 14627, USA; %aquillen@pas.rochester.edu
}  \\
}

\begin{document}
\label{firstpage}
\maketitle

\begin{abstract}
We identify migrating stars in an N-body hybrid simulation of a Milky-Way-like disk.
Outward migration can occur when a star in a low eccentricity orbit lags a short-lived local spiral arm density peak.  
We interpret  short lived local density peaks, that appear and fade on approximately an orbital period,
 as arising from positive interference between spiral density wave patterns that are longer lived.
We find that stars near such a peak can migrate over a significant distance 
in galactocentric radius during the peak lifetime, providing that the peak is sufficiently dense. 
We propose that short lived density peaks, caused by interference between spiral density waves,
 can induce radial migration even when 
there are no spiral density waves that are strong near their corotation resonance.
Using a Gaussian bar model for the potential perturbation associated with a narrow
transient spiral feature, estimates
of the migration rate, angular offset between particle and spiral feature, and maximum eccentricity
for migrators roughly agrees with the values measured in our simulation.
When multiple spiral density waves are present, local density peaks can  appear and
disappear on timescales faster than the timescale estimated for growth and decay of individual waves
and the peak surface density can be larger than for any individual wave.
Consequently, migration induced by transient density peaks may be more pervasive than that
mediated by the growth and decay of  individual patterns and occurring at their corotation resonance.
We discuss interpretation of transient-like behavior in terms of interfering patterns, including
estimating a coherence time for features that appear due to constructive interference, their
effective angular rotation rates and the speed and direction that a density maximum would move
across a galaxy inducing a localized and traveling burst of star formation.
\end{abstract}

\section{Introduction}

%Scattering by spiral structure and molecular clouds heats the stellar disk, moving stars to increasingly 
%eccentric and inclined orbits, and so increases the stellar velocity dispersion \citep{jenkins90,minchev06}.  

Bar and spiral structures can change the mean orbital radius or guiding centers of stars
and so migrate them from their birth radius 
\citep{wielen77,fuchs87,wielen96,sellwood02,roskar08,minchev11,roskar12,grand12,grand12b}. 
The resulting radial mixing is required to account for stellar metallicity distributions \citep{schoenrich09}. 
Many stars, including the Sun, \citep{adams10}, were born in a cluster, \citep{eggen92,desilva07,bubar10} that then dissolved or 
dispersed to other radii in the Galaxy \citep{wielen77,wielen96,simon09,brown10,bland10}.   Understanding migration and heating
mechanisms in the stellar disk is needed to interpret forthcoming surveys of stars in the Galaxy.

Transient spiral structure \citep{sellwood84,toomre90,toomre91,baba09,fujii11,sellwood11}, 
causing both heating and migration, has been modeled
as a diffusive stochastic process \citep{jenkins90,schoenrich09,bland10}.
One difficulty of this approach arises from the small number of Galactic rotation periods in a Hubble time.
The rotation period of the Sun is about 250 Myr, giving only 40 rotation periods during the
lifetime of the Galactic disk.  If spiral arms require a few rotation periods to grow and decay, then only about
a dozen non-interfering patterns can arise during the lifetime of the Galactic disk.  Recent studies
suggest that multiple patterns can co-exist in a galactic disk \citep{henry03,naoz07,meidt09,quillen11,sellwood11}.  
Only if multiple patterns grow and decay in an uncorrelated or non-interfering way can stochastic models
for radial migration be adapted in a straightforward manner to model radial migration.  
Interference between patterns may increase the heating rate \citep{minchev06,minchev12}, induce chaos
in the dynamics \citep{quillen03}, account for
star formation in armlets \citep{henry03}, gaps in local velocity distributions \citep{quillen11},
and modify disk surface brightness profiles \citep{minchev12}.
We focus here on how the presence of multiple spiral density waves influences stellar migration. 

Stellar migration can occur when a star is temporarily trapped in the corotation
region (or resonance) of a spiral density wave \citep{lyndenbell72,sellwood02}.   We outline the physical
mechanism as proposed and described by \citet{sellwood02}.
 The star is captured into the corotation resonance
as the spiral density wave grows.  When there is a fixed rotating spiral perturbation, the Jacobi integral, $E_J = E - L \Omega_s$,
is conserved.  Here $E$ and $L$ are the energy and $z$ component of the star's angular momentum per unit mass,
and $\Omega_s$ is the pattern speed of the spiral density wave.  The star can vary in angular momentum, $L$, while
the spiral density wave is strong.  After the spiral density wave dissipates, the star can be left 
at a different angular momentum, but may not have significantly increased in eccentricity.
A number of numerical studies have
supported this migration mechanism, including the seminal work \citet{sellwood02}, but also \citet{roskar08} using
SPH simulations, and recently \citet{grand12, grand12b} in both N-body and SPH/N-body simulations. 
\citet{grand12,grand12b} illustrated that outward migrating stars lagged a spiral density peak
and vice-versa for inwards migrating stars.  They also showed that migrating stars remained
on nearly circular orbits, as would be expected when $E_J$ is conserved and $\Omega \approx \Omega_s$
during migration.   Here $\Omega$ is the angular rotation rate of the migrating star.
Distributions of the change in  angular momentum as a function of
initial angular momentum, $\Delta L$ vs $L_0$,  
from numerical simulations \citep{sellwood02,roskar12,minchev12}, exhibit features at corotation resonances,  
 with positive variations in $\Delta L$ occurring inside corotation
and vice-versa outside of the resonance.
%(see Figures 7 and 8 by \citealt{roskar12} and  Figures 6 and 7 by \citealt{minchev12}).
The theory (as outlined by \citealt{sellwood02}) for migration due to a corotation  resonance
is simplest in the presence of a single slowly growing and then fading spiral pattern.
We examine here how this mechanism might be modified in the presence of multiple spiral density waves.

In this paper we examine migration in a simulation with multiple spiral density waves.
We study individual
migrating particles to probe the mechanism  accounting for their rapid variations in angular momentum.
We then discuss modifications to the corotation mechanism for radial migration
in a setting with multiple spiral density waves.

\section{Migration in an N-body simulation}

We use the disk galaxy simulation previously described and studied by \citet{quillen11}.
The initial conditions for the simulations were for a model Milky Way galaxy, 
generated with numerical phase phase distribution functions using the method discussed by \citet{widrow08}.
Spectrograms were used to identify a bar and three spiral patterns in this simulation, two with
two arms,  ($m=2$), and one with three arms ($m=3$, see Table 1 by \citealt{quillen11}). 
The spiral structure (shown in log $r$ vs angle) shown in Figure 3 by \citet{quillen11} can be compared with 
similar figures of other studies (see Figure 9  by \citealt{grand12} and 10 by \citealt{grand12b}).   
While $m=2$ and 3 structures
dominate our simulation and that by \citet{grand12b}, the one studied by \citet{grand12} is dominated by patterns with higher numbers
of arms ($m$ between 3--7) and has a flocculent rather than grand-design morphology. 

\begin{figure}
\includegraphics[width=3.5in,trim=0.8in 0.1in 0.5in 0.1in]{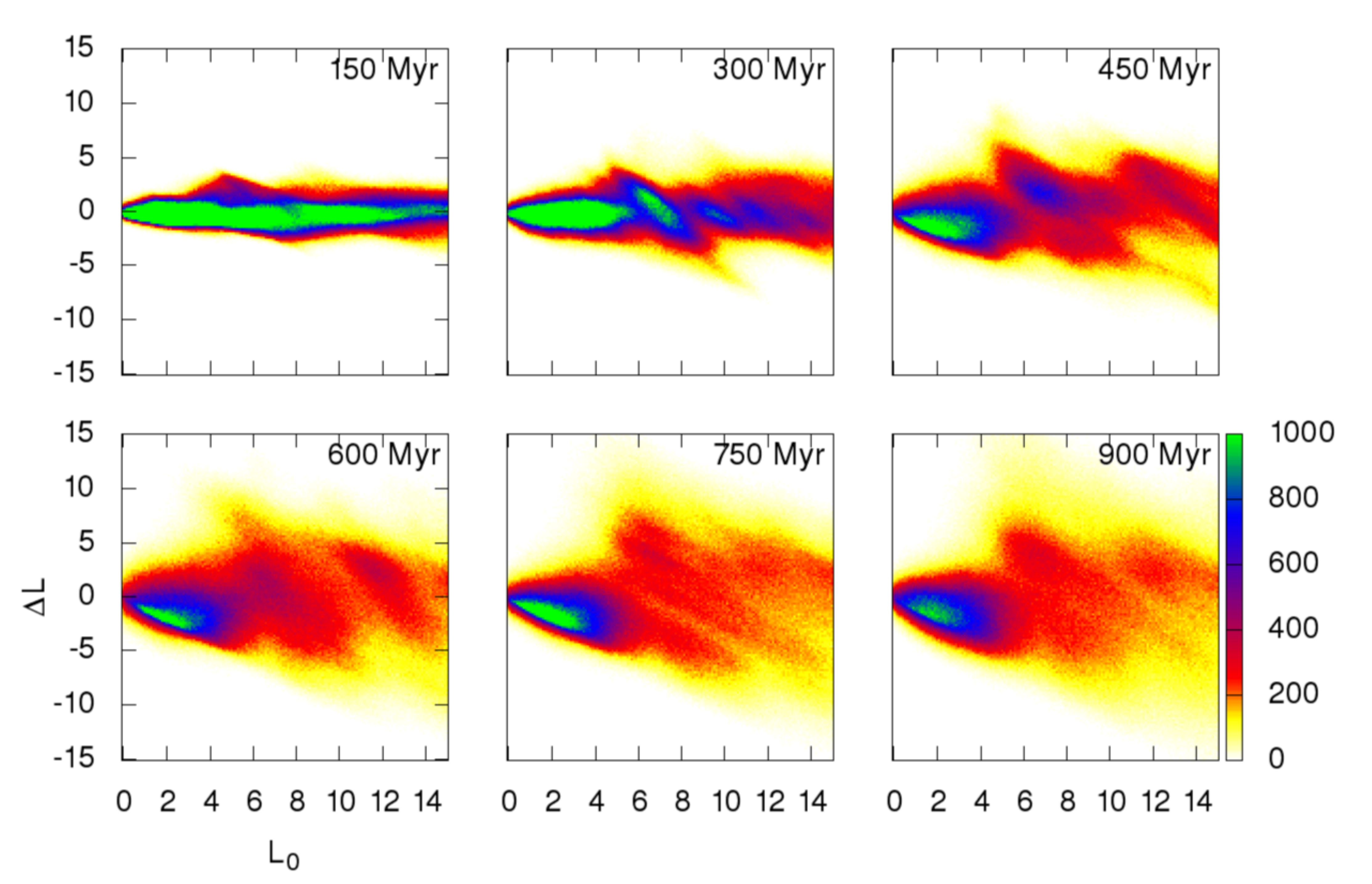}
\caption{
Histograms of the numbers of stars that experienced a 
change in angular momentum, $\Delta L$ (shown on the $y$-axis), 
versus initial angular momentum, $L_0$ ($x$-axis), 
for the disk particles in the simulation.
From left to right the histograms are shown every $0.3$~Gyr. 
Here $L_0$ and $\Delta L$ are  in units of 100 km~s$^{-1}$~kpc.  
}
\label{fig:dL}
\end{figure}

In Figure \ref{fig:dL} we show a histogram of changes in angular momentum, $\Delta L$, computed every 0.15 Gyr,
 as a function of initial angular momentum, $L_0$, in units of 100 km~s$^{-1}$~kpc for the disk particles in the simulation.   
 As have previous studies (e.g., \citealt{roskar12,minchev12}), we see a strong feature associated with the bar's corotation radius at 
 $L_0 \sim 8$ km~s$^{-1}$~kpc, moving progressively outwards as the bar slows down.
 However,  changes in angular momentum occur all over the galaxy and changes in angular momentum persist
 after the bar has grown. This migration  cannot be mediated by the growth and decay of the bar's corotation resonance
 as the bar does not decay.  We conclude that migration is pervasive and on-going.
Because of the number of spiral density waves, this is perhaps not surprising.  
However, the spectrograms (see Figures 4-6 by \citealt{quillen11})
show that the stronger patterns in the simulation contain little power at their corotation radii.  
Only the bar and the $m=3$ pattern contain power near corotation (see Figure 6 by \citealt{quillen11}).
A similar problem was discussed by \citet{grand12}
whose spectrograms showed that most spectral features were strong at
frequencies above or below corotation (see their Figure 6).
Nevertheless they also found that radial migration occurred in their simulation.
\citet{grand12b} illustrated extreme migration near the end of their bar, and we see from the $\Delta L$ vs $L_0$ histograms
that migration in our simulation occurs in the same region.

\subsection{Finding migrators}
\label{sec:finding}

\begin{figure*}
%trim=l b r t 
%
\centering
\mbox{
\subfigure{\includegraphics[width=9.2cm,trim=1.2in -0.2in 0.0in 0.4in]{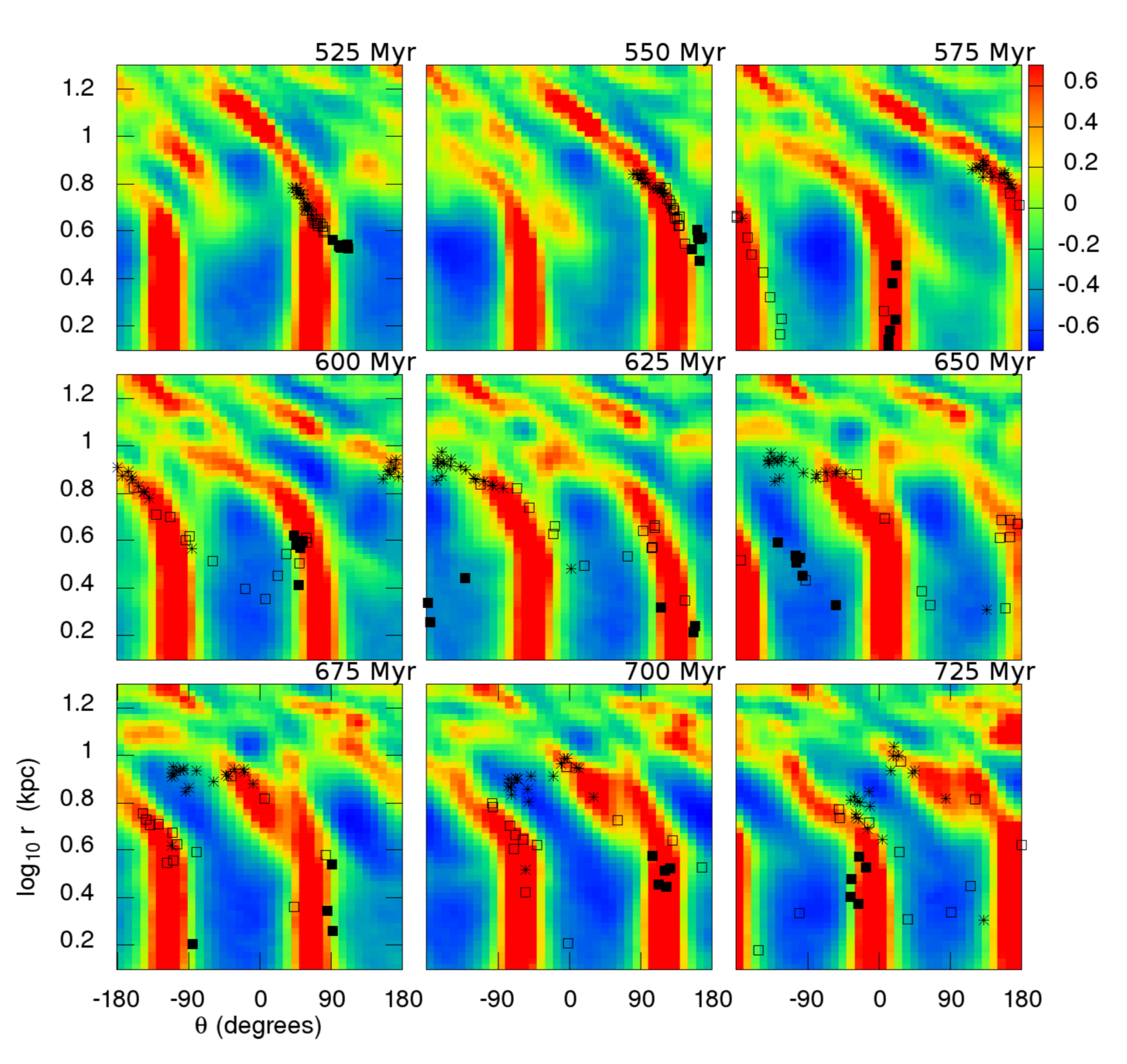}
}\quad
\subfigure{\includegraphics[width=9.8cm, trim=0.5in 0.5in 0.1in 0in]{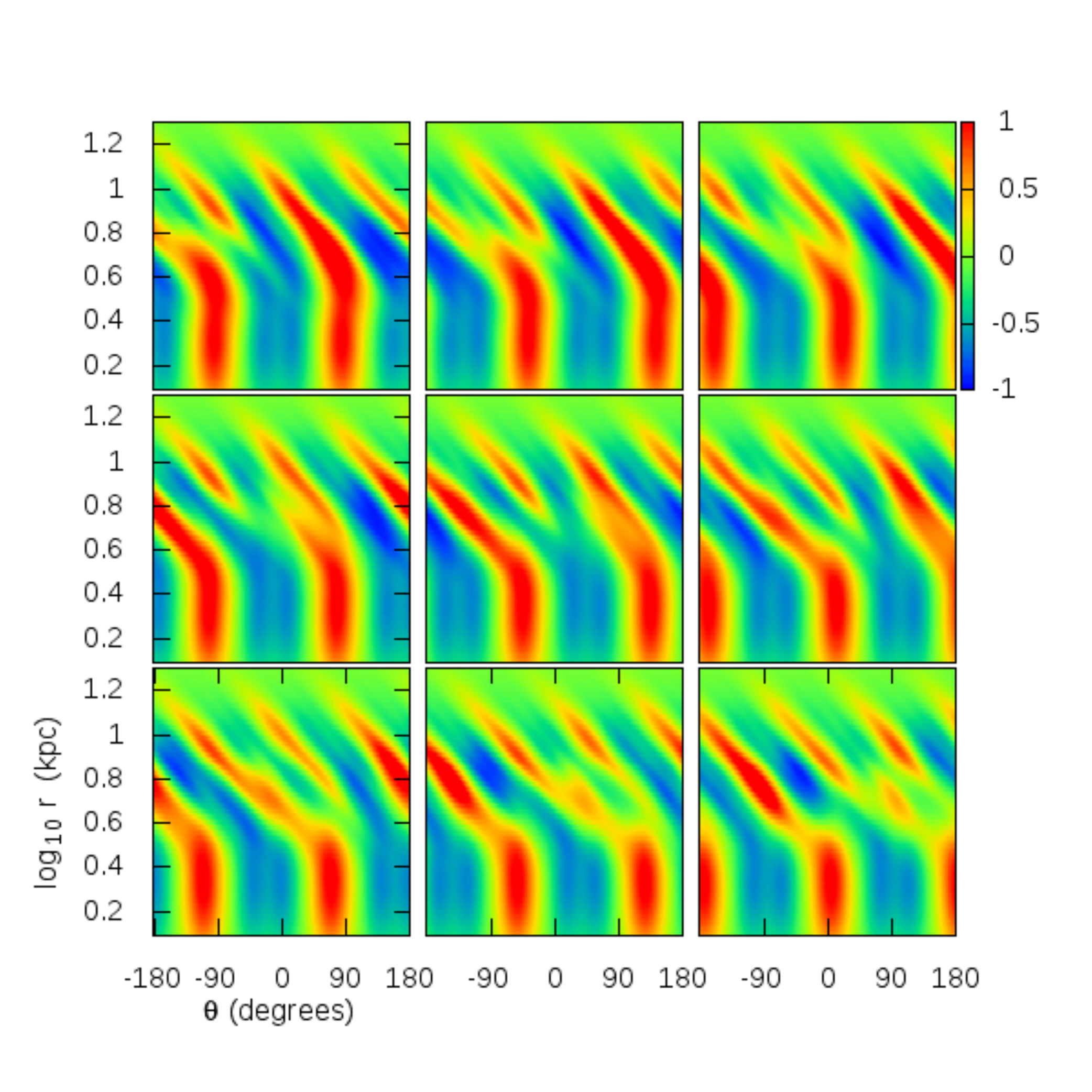}
}}
\caption{
a) Position of 40 stars at different times.  These 40 particles were all within 0.1 kpc of each other
at 500 Myr, just before the top left panel.  Each panel shows a different snapshot
with snapshots plotted every 25 Myr, starting at 525 Myr.
The stars are plotted on top of the surface density distribution 
in cylindrical coordinates.  
The $x$-axis gives azimuthal angle and the $y$-axis log$_{10}$ of the radius in kpc.  
%The stars were coincident in space and angular momentum 
%with a star that migrated outwards along the density peak and is one of the asterisks.  
In the plot, the stars marked with an asterisk have a radial velocity, 
$u \leq -30$, the filled squares $u \geq 30$, and the open squares $|u|< 30$ km~s$^{-1}$.
  The local density peak inducing migration lasted
less than 300 Myr.   Rotation is with positive $d\theta/dt$ so patterns move to the right.  
Low eccentricity stars migrate outwards when they lag (and follow) a local density peak in the disk.
Only particles with small positive radial velocities move outwards.  Particles with higher epicycles 
(the filled and many of the open squares) fail to migrate.   
b)
Surface density distribution similar for the same times as a) but 
for the model described in section \ref{sec:describing} with the parameters listed in 
Table \ref{tab:tab1}.  
%Interference peak distribution and durations for these values are comparable to the conditions in the 
%simulation for the period shown in Figure \ref{fig:movie}a (525 to 700 Myr).
}
\label{fig:movie}
\end{figure*}

We identify particles that migrate large distances in the following manner. 
We first identified particles
with an overall change in angular momentum of $15 < |\Delta L| < 50$ km~s$^{-1}$~kpc.  
Of these `extreme migrators', a randomly selected sample 
were inspected for radial movement in proximity to the bar and spiral arms.  
Of these we chose a single fiducial particle to study in more detail.
In Figure \ref{fig:movie}a we show the trajectory of a particle that migrated from radius $r=4.1$ kpc 
at 500 Myr to 8.6 kpc at 750 Myr, along with 39 other
particles that were near,
(within 0.1 kpc in galactocentric radius and 5$^{\circ}$ in azimuthal angle) 
the chosen particle at $500$ Myrs, 
the start of the chosen particle's outward migration.
The positions of these particles are shown at 525 Myr (just after they are all in the same region)
and at 25 Myr intervals afterwards.
At 500 Myr,
all 40 of these particles have similar angular momentum (within 25 km~s$^{-1}$~kpc of the chosen particle),
however they do not all have the same radial velocity.  We use $u$ to denote the radial velocity component
in galactocentric coordinates with convention positive $u$ for a particle moving toward the galactic center.
In Figure \ref{fig:movie}a
particles with $u \leq -30$ km~s$^{-1}$ are shown as asterisks and include the chosen particle, 
the filled squares are those with $u \geq 30$, and the open squares with $|u| < 30$ km~s$^{-1}$.
Even though the chosen particle is in a nearly circular orbit (and has little epicyclic motion) it has
a negative $u$ corresponding to an outward drift.

%There were 186 particles that were within $0.1$ kpc of the aforementioned particle, 
%%were within an angle $\pm 5^{\circ}$,
%and with angular momentum within 25 km~s$^{-1}$~kpc above or below the chosen particle.

The positions of the 40 particles are  shown in Figure \ref{fig:movie}a 
along with the disk density in cylindrical galactocentric coordinates.
The azimuthally averaged value of the surface density has been subtracted at each radius.  
The $x$-axis is the azimuthal angle, $\theta$, in degrees and the $y$-axis is $\log_{10}$ radius in kpc.
Rotation is in the positive $\theta$ direction (moving to the right in Figure \ref{fig:movie}a) 
and the spiral structures are trailing
with a pitch angle of approximately $24^\circ$ (as measured previously by \citealt{quillen11}).
During the time $500$ to $750$ Myr the chosen particle, and many of the particles
with similar initial velocities, lag just behind a strong local density feature
and it is during this time that the particle moves outwards.  
The local density peak exerts a force in the azimuthal direction on the particle.  This force gives a torque
that causes the particle's angular momentum to increase.
As long as the particle remains approximately the same distance away from
the density peak and lags the peak, it continues to gain angular momentum and moves outwards.

\begin{figure*}
\includegraphics[width=7.5in,trim=0.9in 0.1in 0.1in 0in]{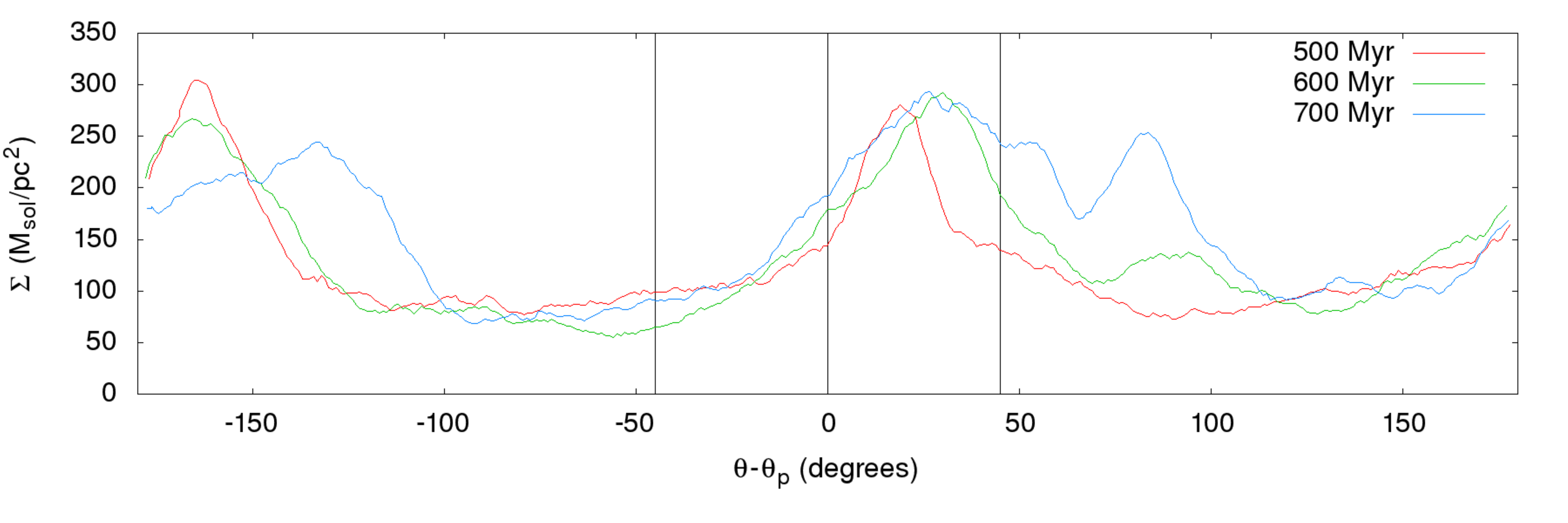}
\caption{
Surface density for our selected particle involved in outward migration (one of the asterisks in Figure \ref{fig:movie}a),  
at three separate times during migration.  The $x$-axis shows the azimuthal angle, $\theta - \theta_p$, 
with respect to the angle of the particle.
Surface density is calculated in an annulus at the particles radius with a width of $\pm 0.1$ kpc and is given in 
units of $M_\odot$~pc$^{-2}$.   Rotation 
is in the positive $\theta$ direction so the particle lags the nearest density peak for the three times shown.  Vertical
lines at $\pm 45^{\circ}$ show region in which we measure the location of the nearest
peak and its density.  The angular difference
$\Delta \theta$ between particle and nearest density peak is shown in Figure \ref{fig:rates}.
The density of the nearest peak and radial migration rate for the same particle
are plotted in Figure \ref{fig:sig_rdot}.
}
\label{fig:peaks}
\end{figure*}

Our fiducial or chosen particle is used in a number of Figures.
Shown in Figure \ref{fig:peaks} is the surface density as a function of angle,
at three different times during the migration for our selected particle.
The angular difference
 between this particle and nearest density peak is shown in Figure \ref{fig:rates}.
The density of the nearest peak and radial migration rate for the same particle
are plotted in Figure \ref{fig:sig_rdot}.

At every timestep we
measured the angular distance between our fiducial particle (one of the asterisks in Figure \ref{fig:movie}a) 
and the nearest density peak.
We denote the angle difference, $\Delta \theta \equiv \theta_{peak} - \theta_p$, 
between the azimuthal angle of the nearest spiral density peak, $\theta_{peak}$, and 
the particle, $\theta_p$; 
We measure $\theta_{peak}$ by 
computing the surface density as a function of azimuthal angle in an annulus with mean radius
equivalent to that of the particle and with radial width $0.2$ kpc.  
Within this annulus we located the highest local maximum disk surface density
that was less than 45$^\circ$ away from the particle, and recorded the angle of this peak.
Shown in Figure \ref{fig:peaks} is the surface density as a function of $\theta- \theta_p$,
at three different times during the migration for our selected particle.
At each time there is a nearby (within 45$^\circ$) density peak.  The particle lags the nearest density
peak and this is consistent with its outwards migration.
Figure \ref{fig:peaks} also shows the surface density in units of $M_\odot$~pc$^{-2}$.  In subsequent discussion
we will use the surface density of the nearest peak to estimate the migration rate of the particle.  The estimated
migration rate will then be compared to the actual migration rate. 

Plotted in Figure \ref{fig:rates}a is 
 the rate of change of angular momentum, $\dot L$, in units of kpc$^{2}$~Myr$^{-2}$, as a function of time, 
 for the selected particle %with trajectory shown in Figure \ref{fig:movie}a, 
 along with the
the  difference between the azimuthal angle of the nearest spiral density peak and the particle, $\Delta \theta$. 
%$\Delta \theta = \theta_{peak} - \theta_p$ where $\theta_p$ is the azimuthal angle of the particle.  
On the bottom panel of the plot we also show the particle's radius as a function of time.
The angular rotation in this simulation has positive $d\theta/dt$ corresponding to counter-clockwise
rotation.   Consequently our plotted angle, $\Delta \theta$, is positive when the particle lags the peak. 
We see from Figure \ref{fig:rates}a that the particle moves outwards ($\dot L$ is positive and radius increases) 
when the particle lags the nearest density peak ($\Delta \theta >0$).  

Our illustration that a particle migrates outward when it lags a density peak is similar and consistent with
the similar illustrations by \citet{grand12,grand12b}.  
The corotation resonance model for migration \citep{sellwood02}  predicts that the particle would be out of
phase with a potential perturbation while it drifts outwards.  In this manner, the offset between particle and density
peak seen in our simulation is consistent with the corotation resonance model for migration.

Figure \ref{fig:rates}a also shows the particle radius as a function of time.  When the particle's
eccentricity is non-zero the particle oscillates radially (known as the epicycle).   
Previous studies have found that corotation resonances are only effective at causing migration
among low eccentricity particles (see Figure 12 by \citealt{minchev12} and associated discussion).
We can see in Figure \ref{fig:rates} that the particle only migrates  when the particle eccentricity is low and that
the eccentricity does not increase during migration.
The positive radial velocity value for migrating particles was due to their outward drift
rather than epicyclic oscillations. 
The corotation resonance model for migration by \citet{sellwood02} is consistent with the
lack of eccentricity variation during migration.
Figure \ref{fig:movie}a shows that particles that have radial velocity $u$ differing by 30--40 km~s$^{-1}$ from
our extreme migrators fail to migrate.  For particles nearly in circular orbits,
eccentricity can be estimated from a maximum $u$ value as $e \sim u/v_c$.
Hence we crudely estimate that only particles with eccentricity $e \lesssim 0.2$ migrate in the vicinity of our
fiducial particle and a maximum migrator eccentricity of $e_{max} \sim 0.2$. 

Figure \ref{fig:rates}b we show a particle from Figure \ref{fig:movie}a
which does not migrate significantly.  
As also seen in the case in which the particle migrated, this particle
has positive $\dot L$ when lagging the nearest density peak, but also negative $\dot L$ when leading the nearest density peak.
As these periods of lagging and leading occur for similar durations and distances to the peak location, 
this particle does not migrate outward like the particle shown in Figure \ref{fig:rates}a.

\subsection{The lifetime of a local density peak that can cause a star to migrate}
\label{sec:lifetime}

\begin{figure}
\includegraphics[width=3.5in, trim=0.8in 0.2in -0.0in 0.3in]{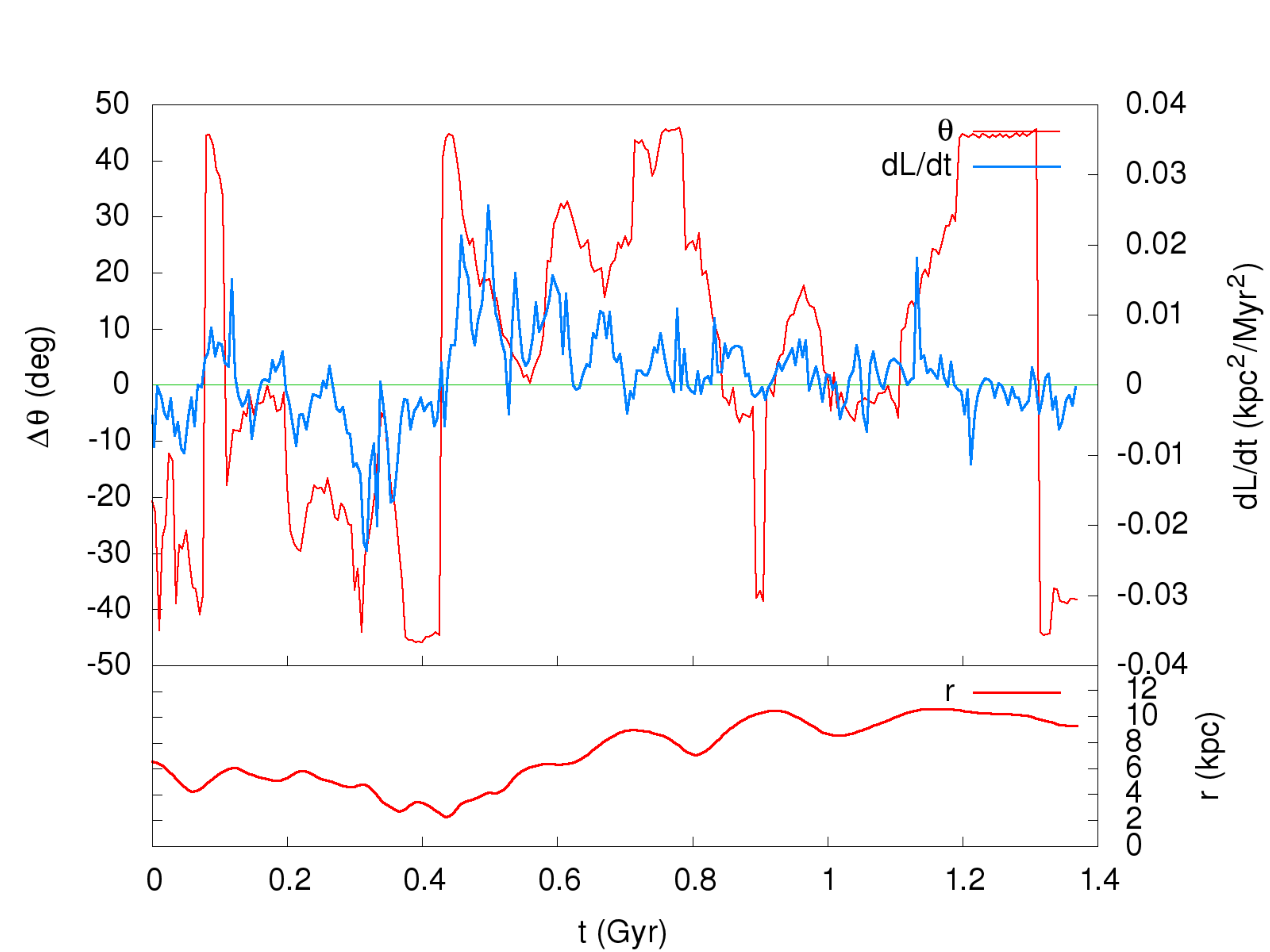}
\includegraphics[width=3.5in, trim=0.8in 0.2in -0.0in 0.0in]{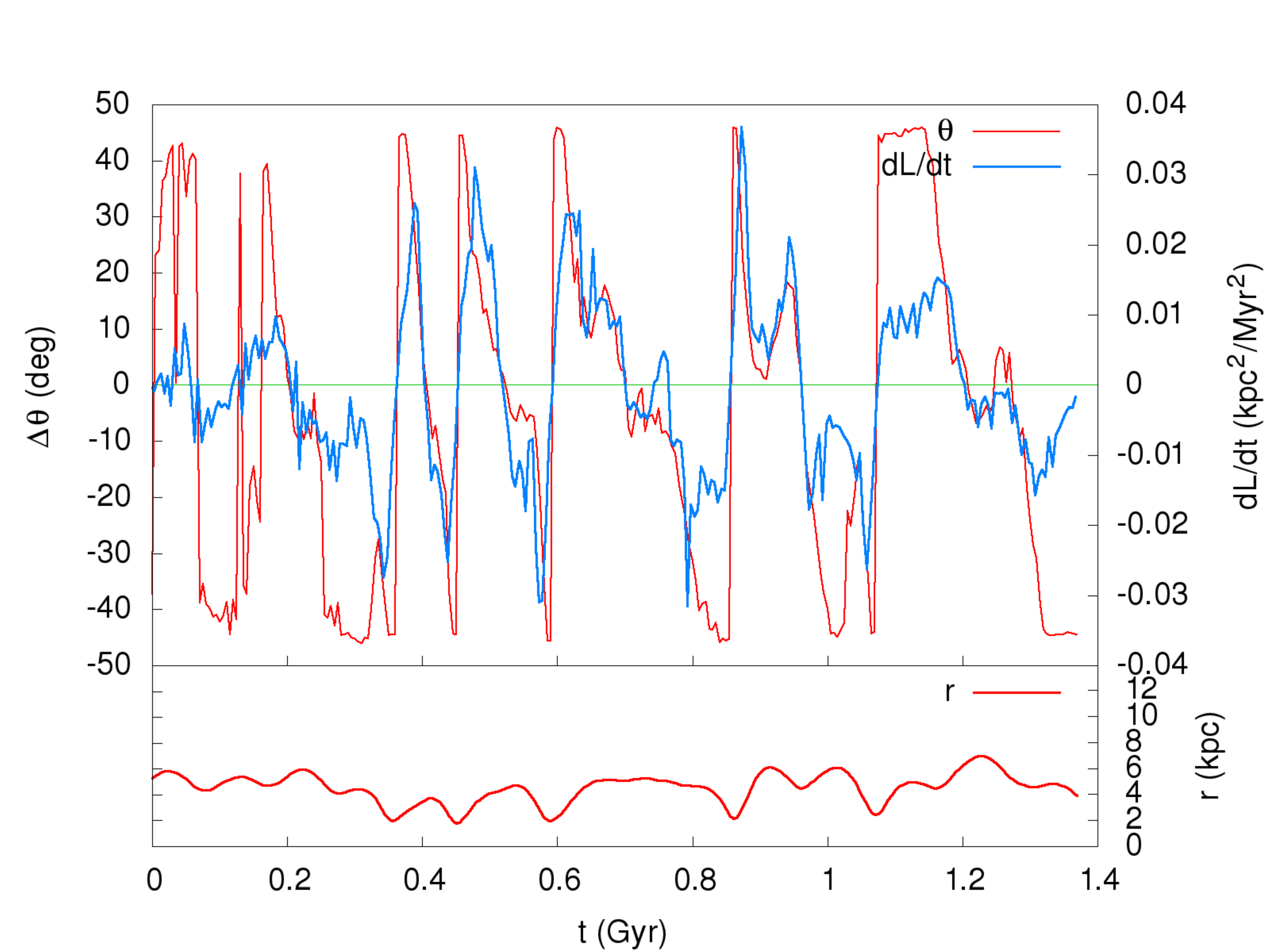}
\caption{
a) For a particle that migrates we plot the rate of change of angular momentum ($\dot L$) 
as a function of time  as a blue line with axis on the right hand side.   
Also   plotted, as  a red line, is the angle to nearest density peak ($\Delta \theta$)
with axis on the left hand side.
Note that outwards migration (positive $\dot L$) only only occurs when the particle lags the density
(positive $\Delta \theta$).   We also show particle radius as a function of time in the lower panel. 
Outwards migration primarily occurs when radial oscillations are not large.
Units for $\dot L$ are kpc$^2$~Myr$^{-2}$.
b) Similar to a) except for a particle that does not migrate a significant amount during the simulation.
}
\label{fig:rates}
\end{figure}

The density feature or peak responsible for the migration of the star shown in Figure \ref{fig:movie}a
appeared at a time of about 475 Myr  and disappeared
at a time of about 750 Myr and so lasted about 275 Myr.
In this simulation the circular velocity is about 200 km~s$^{-1}$ and so the 
rotation period at a Galactocentric radius of 10 kpc is about 300 Myr.  The local peak (causing migration of the particle
with trajectory shown in Figure \ref{fig:movie}a and \ref{fig:rates}a)
 existed less than a rotation period.
 \citet{grand12} followed a single spiral density peak at a radius of about 5 kpc for 120 Myr at which time it disappeared. 
The spiral feature that they traced had a pattern speed (or angular rotation rate) 
that varied with radius and appeared to be
corotating (see their Figure 5).   They demonstrated that it induced radial migration in nearby stars.   
\citet{grand12} commented,
``if there are indeed several wave modes present, it is evident that they must conspire in a specific way in order to produce a spiral arm feature that is Ôapparently corotatingÕ .''  
\citet{grand12} also commented the lifetime of the spiral feature
they followed was short.   
A similar short lived spiral feature was traced by \citet{grand12} with a lifetime of 160 Myr near the end of a bar.
The stars identified by \citet{grand12b} in a barred galaxy simulation with morphology similar to ours also migrated
when they were near a short lived density feature.

\subsection{Transient features caused by interference between patterns}

Swing amplification spiral models (e.g., \citealt{fuchs01}) require a few rotation periods for growth and fading
of the spiral density wave.  However, a density feature that is short lived can be produced when
two spiral density waves interfere.  When density peaks are coherent, a large density peak can be produced from
the sum of the two amplitudes.
This density peak should survive until the two spiral density waves are out of phase.
Consider two spiral density waves with $m_1$ and $m_2$ arms each and patterns speeds, $\Omega_1$ and $\Omega_2$,
respectively.
  In the frame of the first pattern, the other pattern has speed $\Omega_2 - \Omega_1$.
If initially both patterns have maximum at the same angle, after a time  
\begin{equation}
t = { \pi \over m_2 } {1 \over \left|  (\Omega_2 - \Omega_1) \right|},
\end{equation}
the second pattern would  be $180^\circ$ out of phase at the location of the first pattern peak.
We can also consider how long it takes for the second pattern peak to reach the next minimum of the first pattern.
The minimum of these two timescales gives a symmetrical coherence timescale
\begin{equation}
t_{coh} \sim  {\pi \over \max (m_1,m_2) }  {1 \over \left| \Omega_2 - \Omega_1\right|}. \label{eqn:coh}
\end{equation}
Interference peaks are long lived, (long $t_{coh}$), only if the pattern speeds of the interfering
waves are similar.
For patterns with more than one arm, 
 the lifetime of a interference peak is shorter than a rotation period at radius $r_0$
as long as 
\begin{equation}
\left| { \Omega_2 - \Omega_1 \over \Omega_0 }\right| 2 \max(m_1,m_2) \gtrsim {1  }, \label{eqn:cond}
\end{equation}
where $\Omega_0$ is the angular rotation rate of a particle in a circular orbit at  $r_0$. 
We can see if this condition can be satisfied in the vicinity of the end of the bar end
in our simulation.   We take pattern speeds measured previously 
in our simulation (listed in Table 1 by \citealt{quillen11}).
Using a bar angular rotation rate $\Omega_b \approx 40$ km~s$^{-1}$ kpc$^{-1}$  and a two
armed inner pattern with $\Omega_s \approx 30$ (in the same units) and
$\Omega_0  \sim  \Omega_b (r_0/r_b)$ where $r_b$ is the bar's corotation radius,
we estimate 
\begin{equation}
4 \left| {\Omega_b - \Omega_s \over \Omega_0} \right| \sim  {r_0 \over r_b}
\end{equation}
using $m=2$ for both patterns.
Thus condition equation (\ref{eqn:cond}) is satisfied outside bar's corotation radius 
($r_0 > r_b$) for the bar
and the inner two-armed spiral pattern.
The condition for a coherence time shorter than a rotation period
would also be satisfied by a slower three-armed spiral pattern interfering with the bar.

The bar in our simulation has a length of 4 --5 kpc (corresponding to 0.6 --0.7 in log).  
However in the simulation the bar 
appears to  be longer at time 550 Myr than at 700 Myr.   If the feature that caused migration in our simulation
is a result of interference between a bar and a spiral density wave, then we expect both spiral and bar
are in phase during the migration. The addition of spiral waves near the end of the bar might also account
for the some of the variations in morphology near the end of the bar, explaining
why the bar appears to vary in length and how its ends curve at different times. 
At time 500 Myr, the bar extends to 0.8 (in log).   Between 525 and 650 Myr, in the region between
0.6 and 0.9, one end   of the bar progressively curves and becomes more tightly wound.   
Previous studies have also
seen spiral features that appear to become increasingly tightly wound  (e.g., \citealt{grand12b}). 
In other words the winding angle of a density peak appears to tilt as a function of time.
Another way to describe this behavior is with a measure for the angular rotation rate of the density peak that
depends on radius.
A spiral feature that becomes more tightly wound also has a density maximum
 with angular rotation rate that varies with radius. 
Below, we will discuss interpretation of spiral features that wind up (increase in winding angle in time)
or have angular rotation rate that is dependent on radius in terms of interference between steady patterns.

\subsection{Describing the simulation density peaks in terms of interfering waves}
\label{sec:describing}

We can test the possibility that transient structures in our simulation are due to spiral density wave interference
by seeing if we can mimic the appearance of the density perturbations, seen in the simulation,
with a simple interference model.
In the same projection as shown in Figure \ref{fig:movie}a, we have summed three density perturbations
each with a fixed pattern speed.
Each perturbation is described as a surface density perturbation with a Gaussian amplitude
\begin{equation}
\Sigma(r, \theta,t) = \Sigma_i \exp \left( -{(\log_{10}r - l_{i})^2 \over 2 w_i^2}\right)
f (\phi_i (r, \theta, t)) \label{eqn:pat1}
\end{equation}
with function
\begin{equation}
f(\phi) = \cos(\phi) + 0.25 \cos(2 \phi). \label{eqn:pat2}
\end{equation}
For each pattern the parameter $\Sigma_i$ describes the density amplitude.  
The function, $f(\phi)$, is somewhat more peaked at $\phi=0$
than a cosine and allows sharp features to be present in the surface density.  
The amplitude peaks at a radius $r_i$ such that $\log_{10} r_i = l_i$, consequently $l_i$ is
a parameter that describes the mean of the radial region in log space that contains the wave.
The amplitude extends a radial range controlled by the parameter $w_i$.
We define a spiral argument for the $i$-th spiral density wave
\begin{eqnarray}
\phi_i(r,\theta,t) \equiv \alpha_i \ln (r/r_b) - m_i (\theta - \Omega_i (t-t_0) - \theta_i). \label{eqn:pat3}
\end{eqnarray}
The number of arms or azimuthally symmetry (for a bar) is set by $m_i$.
The pattern speed is set by $\Omega_i$ and $\alpha_i$ controls the winding angle.
The bar perturbation has $\alpha_i=0$ but when $\alpha \ne 0$,  the spiral perturbations are logarithmic.
A logarithmic spiral density wave with $m_i$ arms can approximately be described as having 
surface density and potential dependent on the angle $\phi_i$ and with amplitude slowly varying with radius.
The offset of each pattern at time $t=t_0$ is set by a radius, $r_b$, and angle $\theta_i$.
Here the winding angle, $\gamma_i$,   depends on $r d\theta/dr$ in a density peak with 
\begin{equation}
\cot \gamma_i  = \beta_i = {\alpha_i \over m_i}. \label{eqn:winding}
\end{equation}

A model that contains two spiral patterns and a bar is shown in Figure \ref{fig:movie}b with parameters
listed in Table \ref{tab:tab1}.  The listed parameters are for each pattern: a pattern speed ($\Omega_i$),
number of arms ($m_i$), width parameter ($w_i$),  winding parameter ($\alpha_i$), log radius
of maximum ($l_i$) and angular offset ($\theta_i$).
Figure \ref{fig:movie}b, showing the model, can be compared to Figure \ref{fig:movie}a, showing the simulation.
We have assumed rotation with $d\theta/dt=0$, and trailing spiral patterns corresponding to negative $\alpha_i$.
The pattern speeds listed in Table \ref{tab:tab1} are given in terms of the bar's pattern speed
and the offset angles given in terms of the bar's orientation at $t_0 = 550$ Myr, corresponding
to the top left panels in Figures \ref{fig:movie}a and b.
The patterns speeds, $\Omega_i$, are given in units of the bar's
angular rotation rate that is $2.3^\circ$ Myr$^{-1}$ (equivalent to the 0.040 radians Myr$^{-1}$ 
or 40 km~s$^{-1}$~kpc$^{-1}$ measured
 from a spectrogram and listed in Table 1 by \citealt{quillen11}); (to convert from radians Myr$^{-1}$ to 
km~s$^{-1}$~kpc$^{-1}$ multiply by 1000).
%The bar's corotation radius is at about 5 kpc (with log of 0.7).
At early times,  (top panels in Figure \ref{fig:movie}b) the spiral features reasonably well match the simulation
with  morphology shown in Figure \ref{fig:movie}a.  However, at later times, the model and
simulation are increasingly different.
The model is likely too simplistic.  To better match the simulation
we would probably require 
a model containing additional patterns, 
and with patterns that are not fixed in time but slowly vary 
as the bar slows down. 
 
The two-armed structure in the model
has an angular rotation rate of 34.4  and the three armed one
has 28 km~s$^{-1}$~kpc$^{-1}$.  These patterns were not listed in Table 1 by \citealt{quillen11}
as prominent patterns in the simulation.  However they do correspond to fast features seen
in the spectrograms. The three armed structure corresponds to a peak at about 5~kpc at an angular
frequency of 0.090 radians Myr$^{-1}$ shown in the spectrogram of Figure 6 by \citet{quillen11}
(angular frequency is $m$ times the pattern speed).
The two armed one corresponds to a fainter feature just past the end of the bar 
at an angular frequency of 0.07~Myr$^{-1}$ present in the spectrogram shown in see Figure 4 by \citet{quillen11}, and that may be 
associated with a clearer feature in the $m=4$ spectrogram (their Figure 5) at an
angular frequency of about 0.14~Myr$^{-1}$.  %These patterns are very close to their corotation radii.  

%Figure \ref{fig:movie}b resembles Figure \ref{fig:movie}a but is consistent with the pattern speeds
%that are seen as peaks in the spectrograms.  
The reasonable correspondence between the model
Figure \ref{fig:movie}b and the simulation in Figure \ref{fig:movie}a
the two figures suggests that an interference model can reproduce
transient spiral features in this simulation.
The feature causing migration of the star shown in Figures \ref{fig:movie}a 
(and crudely modeled in Figure \ref{fig:movie}b) 
likely appears and disappears due
to interference between the fast two and three-armed structures with patterns listed in Table \ref{tab:tab1}
and are consistent with faint peaks seen in the spectrograms.

Using the the bar and two-armed spiral pattern speed (and $m=2$ for both) listed in Table \ref{tab:tab1} 
and equation (\ref{eqn:coh}) the coherence timescale for peaks arising due
to positive interference is approximately 1.8 bar rotation periods where 
a bar rotation period is approximately 
160 Myr.
Similarly the coherence time between bar and three-armed and two and three armed structures is
0.6 and 1 bar rotation periods.   These short coherence timescales are consistent both with the timescale
we estimated for the lifetime of the peak causing stellar migration and discussed previously 
in section \ref{sec:lifetime}.
%and the  correspondence between the simulation and model density distributions shown 
%in Figure \ref{fig:movie}.

\begin{table}
\vbox to 85mm{\vfil
\caption{\large Density Wave Model \label{tab:tab1}}
\begin{tabular}{@{}lccrccr}
\hline
Pattern            &  $\Omega_i$ &  $m_i$ & $\alpha_i$& $l_i$ & $w_i$ & $\theta_i$ \\ \hline
Bar                  &     1.00             &   2      &   0           &   0.33   &  0.23  &  0$^\circ$ \\
Two-armed     &     0.86            &   2       &   -3         &   0.71   & 0.14   &  -10$^\circ$ \\
Three-armed  &     0.70            &    3      &   -6          &   0.90  &  0.17  &  10$^\circ$ \\
\hline
\end{tabular} 
\medskip 
{\\ A three pattern density wave model with parameters chosen to match
the simulation at around 500 Myr.  
Here angular rotation rates for each pattern, $\Omega_i$ are given in units of the bar's
angular rotation rate. 
%that is $2.3^\circ$ Myr$^{-1}$
% (equivalent to the 0.04 radians Myr$^{-1}$ measured with from a spectrogram and listed in Table 1 by \citealt{quillen11}).  
The number of arms or symmetry in the case of the bar is given by $m$.  The winding angle
is determined from $\alpha_i$ and $m_i$ using equation (\ref{eqn:winding}).  The density peak
is at a radius such that $l_i = \log_{10} r$.  Each pattern extends over a radial range set by $w_i$.
Each pattern is described by Equations \ref{eqn:pat1} - \ref{eqn:pat3} and using $\log_{10} r_b = 0.6$
with $r_b$ in kpc.
The amplitude of each structure has $\Sigma_i = 0.7$ in units of the azimuthally averaged value.
Here the offset angle, $\theta_i$, 
is given in terms of that for the bar that has major axis at $130^\circ$ at time $t_0 = 550$ Myr, corresponding
to the second panel in Figure \ref{fig:movie}a.
The resulting density is shown at different times in Figure \ref{fig:movie}b.  Parameters
were adjusted so that the density field is similar to that seen in the simulation and shown in 
 Figure \ref{fig:movie}a.
}
 \vfil}
\end{table}

\section{Properties of peaks arising from interference between two spiral density waves} 

The corotation model for migration is consistent with the low eccentricities of migrating particles 
and that migration primarily takes place when a particle lags a density peak.   However 
our spectrograms don't show patterns (other than the bar) that are strong in their
corotation region.  Furthermore the peaks that we see in our simulation causing migration survive less than
rotation period and so are likely caused by interference between spiral density waves.  Spiral patterns need not
be long lived, however it is unlikely that spiral patterns appear and decay on timescales 
less than a rotation period.   We consider the possibility that local peaks, 
caused by positive interference between two patterns, can cause stars to migrate.    
We first consider properties of local density peaks produced by interference.  We then  
modify the corotation resonance model for migration in this setting.

\subsection{Angular rotation rate of an interference peak}

Using a model for two density waves we now compute the angular rotation rate of a local density peak caused by
positive interference between the two waves.  
Spiral density waves are often described in terms of the number of arms, an amplitude (that is a function
of radius), a winding
angle and a pattern speed.    Except for the number of arms, each of these parameters could vary in time, and their radial 
dependence could also vary with time. This is a large and poorly constrained multidimensional parameter space
for variations.  
%Swing amplification models predict amplitude and winding angle variations on timescales of order a few rotation periods (e.g., \citealt{fuchs01}).
%
To separate between variations caused by single spiral patterns that vary in time, and phenomena caused
by interference we can consider two fixed patterns that have different pattern speeds.
We use  a spiral argument for the $i$-th spiral density wave (similar to equation \ref{eqn:pat3})
\begin{eqnarray}
\phi_i(r,\theta,t) \equiv \alpha_i \ln r - m_i (\theta - \Omega_i t - \theta_{i}) \label{eqn:phii}
\end{eqnarray}
where we have dropped the constant $r_b$ and the angle  $\theta_{i}$ describes an angular offset at time $t=0$.
With  two logarithmic spiral perturbations  the mass surface density as a function of time
\begin{equation}
\Sigma(r,\theta,t) = A_1(r) \cos(\phi_1) + A_2(r) \cos(\phi_2) \label{eqn:sig}
\end{equation}
%Here $m_i,\Omega_i$ are number of arms and pattern speed for the $i$-th pattern and $\alpha_i$ determines the winding angle.  
The amplitude functions $A_i(r)$ we assume are slowly varying with  radius.

Consider a location and time where there is a local density peak so that $\Sigma$ has
an extremum with ${\partial \Sigma \over \partial \theta}(r_0,\theta_0,t_0) =0$.
By taking the derivative of equation (\ref{eqn:sig})
the extremum occurs where
\begin{displaymath} 
A_1(r_0) m_1 \sin(\phi_1(r_0,\theta_0,t_0) + A_2(r_0) m_2 \sin(\phi_2(r_0,\theta_0,t_0)) = 0. 
\end{displaymath}
We consider small perturbations in time and angle about this moment and position; $t=t_0 + dt$ and
$\theta = \theta_0 + d\theta$ and require that this new position and time is also a local density peak 
and so is also an extremum so that ${\partial \Sigma\over d \theta}(r_0, \theta_0 + d \theta, t_0 + dt) =0$.
By expanding this partial derivative to first order in $dt$ and $d\theta$ we can estimate $d\theta/dt$.
To first order the density peak moves with angular rotation rate
\begin{eqnarray}
\Omega_p = 
{d\theta \over dt} &\approx& {m_1 A_1 \cos \phi_1 \Omega_1 + m_2 A_2 \cos \phi_2 \Omega_2 \over m_1 A_1 \cos \phi_1 + m_2 A_2 \cos \phi_2}  \nonumber \\
&\sim& {m_1 A_1 \Omega_1 + m_2 A_2 \Omega_2 \over m_1 A_1 + m_2 A_2} \label{eqn:dthetadt}
\end{eqnarray}
where in the second step we have assumed that the local density peak is near the density maximum of both waves
where $\phi_1 \sim \phi_2 \sim 0$.

The above expression describes the angular rotation rate of the density peak.  We see that 
its angular rotation rate lies between the the pattern speeds of the two density waves, with value depending upon the amplitudes of
each wave.  Consider a situation with the first wave faster than the second one, $\Omega_1 > \Omega_2$,
and with peak amplitude at a smaller radius than the second one.   In this case $A_1$ and $A_2$ are smoothly varying
with radius and $A_1$ is high at small radii and $A_2$ high at larger radii.
At an inner radius where the first wave dominates,
equation (\ref{eqn:dthetadt}) implies that $\Omega_p$ is near $\Omega_1$ but at a radius where the second pattern
dominates $\Omega_p$ would be near $\Omega_2$.  The peak would effectively vary in pattern speed with pattern
speed decreasing as a function of increasing radius.  Density peaks with an angular rotation rate that depend on radius
 have been identified in simulations, for example
see Figure 5 by \citet{grand12} and Figure 8 by \citet{grand12b}.    
The maximum angular rotation rate is that of the faster pattern, and this is consistent with the angular rotation
rate of the feature identified by \citet{grand12b} approaching that of the bar at small radius (see their Figure 8).
The variation in $\Omega_p$ with radius causes the winding angle of the
density feature to vary in time or wind up (as shown in figure 7 by \citealt{grand12b}).
We note that other types of transient models might also predict such behavior, for 
example swing amplification models predict that the winding angle varies in time as the amplitude varies.  
A swing amplified model would also effectively give a peak with angular rotation rate that varies with radius.

The above expression for the angular rotation rate (equation \ref{eqn:dthetadt}) does not depend upon $\alpha_i$
or the winding angles of the patterns.   Consequently it can be used for $\alpha_i=0$ and so can describe a bar-like
perturbation that has a fixed angle as a function of radius.  We can consider interference between a fast bar perturbation
and a slow local spiral perturbation.   Outside a bar's corotation radius, the bar pattern speed would be faster
than the local angular rotation rate.  Spiral patterns are often observed in simulations to be strong inside their corotation
radius (or at radii where the pattern speed is larger than the local angular rotation rate; e.g., \citealt{quillen11}).
In a setting where a local spiral pattern interferes with perturbations from a bar, we would expect that 
the effective angular
rotation rate $d\theta/dt$ of an interference peak between would lie between the bar's and the spiral's pattern
and ranging at small radius with an angular rotation near that of the bar to near that of the spiral at larger radius.
In this setting the effective angular rotation rate of the interference peak would necessarily pass a point where
$d\theta/dt$ is equal to the local angular rotation rate and so the interference peak would act as if it were
going through its corotation resonance.  This is likely the situation in our simulation and causing migration.
 Bar corotation regions are seen as location where migration is common
even after the formation of the bar (in our Figure \ref{fig:dL}, but also see Figure 7 by \citealt{minchev12}).
However slow spiral structure in the vicinity of a bar would cause interference peaks with nearly corotating angular rotation
rates.  This may in part account for the efficacy of bar corotation regions in causing stellar migration.

\subsection{Radial position of the maximum density}

Above we considered the azimuthal location of peaks. We now consider the radial location of  the maximum density
in a local peak and how it would vary in time.  In our simulation we see local peaks usually appear at a small radius and move
outwards, but sometimes appear at a large radius and move inwards.   We can try to understand the speed and direction
of motion using the same model as we used above.   A localized burst of star formation
could be caused from density wave interference.   
As long as the amplitudes are only slowly varying in time and radius then the local density primarily depends on the arguments.
When the two patterns lie directly on top of one another then the density  is a maximum.
Consider a radius, angle and time where the two waves constructively add or $\phi_1(r_0,\theta_0,t_0) = \phi_2(r_0,\theta_0,t_0)=0$.
Consider a nearby radius $r_0+dr$ where the two waves also constructively add but at time $t_0 + dt$
so that $\phi_1(r_0+dr,\theta_0+d\theta,t_0+dt) = \phi_2(r_0+dr,\theta_0+d\theta,t_0+dt)=0$.
We can solve for $\theta_0 + d\theta$ using equation (\ref{eqn:phii}) for both angles, 
\begin{eqnarray}
\theta_0 + d\theta &=&  {\alpha_1 \over m_1}\ln (r_0 + dr)  + \Omega_1 (t_0 + dt)   \nonumber \\
&= &  {\alpha_2 \over m_2}\ln (r_0 + dr)  + \Omega_2(  t_0 + dt) . 
\end{eqnarray}
Expanding to first order and relating $dr$ to $dt$, we find that the density maximum moves inwards or outwards in radius at a speed
\begin{equation}
\left. {dr \over dt}\right|_{peak} = {r_0 (\Omega_2 - \Omega_1) \over \alpha_1/m_1 - \alpha_2/m_2}
 ={r_0 (\Omega_2 - \Omega_1) \over \beta_1 - \beta_2}, \label{eqn:rpeak}
\end{equation}
where we have defined a parameter
\begin{equation}
\beta_i \equiv {\alpha_i \over m_i} =  \cot \gamma_i. \end{equation}
Here larger $\beta_1, \beta_2$ correspond to more tightly wound structures.

We can see from equation (\ref{eqn:rpeak}) that
the peak maximum moves slowly if the two pattern speeds are similar, and this is expected as they would
remain in phase for longer. 
If we adopt $\Omega_1 > \Omega_2$ then 
the direction of motion depends on the sign of the denominator and this depends on the winding angles
of each pattern (equation \ref{eqn:winding}).  If the faster pattern is more tightly wound
than the other ($\beta_1 > \beta_2$), then the density maximum would move inwards, otherwise it would move outwards.
In the first case we would expect a burst of star formation that progressively moves inwards in radius,
and in the second case moving outwards in radius.  Star bursts propagating inwards would occur if the center of the galaxy
had more open spiral structure than the outer regions of the galaxy.

The denominator of equation (\ref{eqn:rpeak}) is only zero when the two waves have the same winding angle
and so the two patterns are  in phase at all radii simultaneously.  
If two patterns have nearly the same winding angles then the denominator is large.  This implies
that the peak is nearly in phase at all radii simultaneously.  We would expect a burst of
star formation across the density peak that is nearly coeval at all radii.
We note that the above expression neglected variation of the amplitudes $A_1,A_2$ with radius and time, but
expect these would not strongly affect the drift rate, $\dot r$, 
as long as the density variations are more strongly dependent on 
angle than radius.

It may be useful to put equation (\ref{eqn:rpeak}) in units of pc/Myr
\begin{equation}
 {\dot r }_{peak} = 200 ~{\rm pc~Myr}^{-1}
{(\Omega_2 - \Omega_1) \over \Omega_0(\beta_1 - \beta_2)} \left({v_c \over 200 {\rm km~s}^{-1}} \right). 
\end{equation}

There may be  a progression of ages of star clusters and moving groups with stars closer to the Galactic center
being younger in the Sco Cen region (Eric Mamajek, private communication).  
Such transient peaks would be expected within an interference model
 if the outer galaxy is more tightly wound than the inner galaxy.

\section{Migration by local density peaks}

Above we have illustrated how interfering spiral density patterns can cause a local density enhancement to appear for
a short period of time.    We now consider whether such a peak can induce stars to migrate.
We first review a Hamiltonian model for migration caused by a corotation resonance based on
the seminal work by \citet{sellwood02}. 

\subsection{Corotation resonance model for migration}

For an axisymmetric system we can describe orbits in terms of a Hamiltonian that is a function of the angular momentum, $L$, and
the momentum associated with epicyclic motion, $J$.  
These two momenta are conjugate to the an angle $\theta$ that is approximately the azimuthal
angle in the plane, and the epicyclic angle, $\varphi$;
\begin{equation}
H_0(L,J; \theta, \varphi) = g_0(L,J) \sim g_0(L) + g_1(L) J + g_2(L) {J^2 \over 2} .
\end{equation}
As the system is axisymmetric, the Hamiltonian does not depend on the two angles, so 
$H_0, J$ and $ L$ are all constants of motion.
Above and on the right we have expanded the unperturbed Hamiltonian to second order in $J$.
The angular rotation rate of a particle in a circular orbit in the unperturbed potential, $\Omega(L)$, is
\begin{equation} 
\Omega(L) = { \partial g_0(L) \over \partial L}. 
\end{equation}
The epicyclic frequency  $\kappa(L) = g_1(L)$.
For a flat rotation curve $g_0(L) = v_c^2 \ln L$,  $\Omega(L) = {v_c^2\over L}$ 
and $\kappa(L) = \sqrt{2} \Omega(L)$.

We can consider a logarithmic perturbation to the gravitational potential in the form
\begin{equation}
V(r, \theta,t) = \epsilon f(\beta \ln r - (\theta-\Omega_s t)) \label{eqn:vf}
\end{equation}
 where $r$ is the radius, the  perturbation has amplitude $\epsilon(t)$, 
pattern speed $\Omega_s$, and  winding angle $\gamma$,
with $\cot \gamma = \beta$.   
We  consider a system with $\theta$ increasing in the direction of rotation.  
In this case a trailing logarithmic spiral has $\beta <0$.
The above perturbation depends on radius rather than our action angle variables.
If we take the low eccentricity limit and average over the dependence of $r$ on the epicyclic angle then
$r \sim L/v_c$ where $v_c$ is the circular velocity with angular momentum $L$.
In the vicinity of a corotation resonance (and distant from any Lindblad resonances and in the low eccentricity limit)
the Hamiltonian depends on angular momentum
$L$ and azimuthal angle $\theta$ alone;
\begin{equation}
H(L,\theta)  = g_0(L) + \epsilon(t) f(\beta \ln L -  (\theta - \Omega_s t)). \label{eqn:HL}
\end{equation}
where we have replaced $r$ with $L$ and removed an arbitrary constants from $\theta$ and $t$.
\citet{sellwood02} have illustrated that in the vicinity of a corotation resonance, the epicyclic amplitude
is not affected.

%The above Hamiltonian is appropriate in the low eccentricity limit and be found by averaging a more
%general Hamiltonian (e.g., that introduced by \citealt{cont75}) over  the epicyclic angle.
%The second term in equation \ref{eqn:HL}, $\epsilon(t)f(\beta \ln L -  (\phi - \Omega_s t))$,

It is useful to define an angle 
\begin{equation}
\phi \equiv \beta \ln L - (\theta - \Omega_s t ). \label{eqn:phi}
\end{equation}
with  time derivative
\begin{equation}
\dot \phi = \beta {\dot L \over L} - (\dot \theta - \Omega_s).
\end{equation}
The potential perturbation function, $f(\phi)$, may be periodic.
For example the function $f(\phi) = \cos (m \phi)$ gives an $m$ armed spiral structure.
This type of perturbation is equivalent to the model discussed by \citet{sellwood02} and shown in their Figure 6. 
Previously we discussed density perturbations using a similar form
for the perturbations.   Here we focus on perturbations to the gravitational potential.
%Wrong!  We can chose $\theta,t$ such that $\phi=0$ near a peak in density, corresponding to
%a minimum in the gravitational potential where we expect $\epsilon f(0) <0$.
%A particle lagging the potential minimum has small $\theta <0$.

Hamilton's equations (using the Hamiltonian of equation \ref{eqn:HL}) are
\begin{eqnarray}
\dot L &=& -{\partial H \over \partial \theta} = \epsilon f'(\phi)\label{eqn:dotL} \\
\dot \theta &=& {\partial H \over \partial L} = \Omega(L) + \epsilon f'(\phi) {\beta \over L} .
\end{eqnarray}
Inserting $\dot\theta$ and $\dot L$ into the expression for $\dot \phi$ gives
\begin{eqnarray}
\dot \phi &= & \Omega_s - \Omega(L).\label{eqn:dotp}
\end{eqnarray}
The above equation implies that when $\Omega(L) \sim \Omega_s$ (in the vicinity of corotation), 
the argument $\phi$ is nearly constant
and the angular momentum  increases or decreases depending upon the sign on
the right hand side of equation (\ref{eqn:dotL}).
As $\epsilon f(\phi)$ is the perturbation to the gravitational potential we can also write equation (\ref{eqn:dotL}) 
(one of Hamilton's equations) in the following way
\begin{equation}
\dot L = -{\partial \Phi \over \partial \theta} = - ({\bf r} \times \nabla \Phi) \cdot {\bf \hat z}
\end{equation}
where $\Phi$ is the entire gravitational potential.   The angular momentum drift, $\dot L$, arises from the torque from the 
non-axisymmetric portion of the potential perturbation.  Lagging the potential minimum we expect $\epsilon f'(\phi) >0$ 
corresponding to $\dot L>0$.  Particles lagging a potential minimum would migrate outwards.

A maximum migration rate is
\begin{equation}
\dot L_{max} \sim   \max |\epsilon f'(\phi)| \label{eqn:Lmax}
\end{equation}
leading to a maximum radial migration rate of order 
\begin{equation}
\dot r_{max} \sim { \dot L_{max} \over v_c}. \label{eqn:rmax}
\end{equation}
Note that the maximum torque on a particle can be directly estimated from the maximum tangential force
at a given radius.
%A parameter $Q_T$ has been used to describe  the maximum tangential force in a galaxy disk divided by the mean radial force (e.g., \citealt{combes81}) and is defined in an annulus with radius $r$.   
%For a flat rotation curve 
%\begin{equation} Q_T(r) \equiv \left. {\partial \Phi(r,\theta) \over \partial \theta} \right|_{max} {1 \over v_c^2} \end{equation}
 %where the maximum value is taken among different values of $\theta$ but fixing $r$.

In the rotating frame with $\theta' = \theta - \Omega_s t$, the Hamiltonian becomes
\begin{equation}
K(L,\theta')  = g_0(L) -L \Omega_s + \epsilon(t) f(\beta \ln L -  \theta'). 
\end{equation}
As $K$ is independent of time, it is conserved and known as the Jacobi integral.
Expanding this Hamiltonian about the angular momentum corresponding to corotation 
or $L = L_s + J$, with  $\Omega(L_s) = \Omega_s$, 
\begin{equation}
K(J,\theta') =  {\rm const} -  \Omega'(L_s) {J^2 \over 2} + \epsilon f(\beta \ln L_s - \theta') + .... \label{eqn:KK}
\end{equation}
with $\Omega'(L)  = -v_c^2/L^2$ for a flat rotation curve.
This gives an equivalent description of the migration model in the corotating frame.
Up to this point we have not assumed any particular form for the potential perturbation,
only a pattern speed.  Nevertheless the maximum migration rate can be estimated
from the maximum strength of the derivative of the potential perturbation
(equations \ref{eqn:Lmax}, \ref{eqn:rmax}).

\subsection{Corotation resonant width and eccentricity limit}

In the WKB approximation a logarithmic spiral density perturbation  comprised of a single Fourier component
\begin{equation}
\Sigma (r, \theta,t) = -S_m \Sigma_0 \cos (\alpha \ln r - m(\theta - \Omega_s t))  \label{eqn:sigmacos}
\end{equation}
has gravitational potential 
\begin{equation}
V(r,\theta,t) \approx { 2 \pi G S_m \Sigma_0 \over |\alpha| r}\cos (\alpha \ln r - m(\theta - \Omega_s t)), \label{eqn:vcos}
\end{equation}
(computed with wavevector $k \approx \alpha/r$ as by \citealt{B+T}).  Here $S_m \Sigma_0$ is the  
strength of the $m$-th Fourier coefficient of the surface density.  $\Sigma_0$ is the azimuthally averaged
density and $S_m$ is an amplitude in units of $\Sigma_0$.
Previously we have used a winding parameter $\beta = \alpha/m$ so that the argument
can be written $\alpha \ln r - m(\theta -\Omega_s t) = m[\beta \ln r - (\theta-\Omega_s t)] = m \phi$.
In this case  the function $f$ of Hamiltonian equation (\ref{eqn:HL}) is a cosine, and the perturbation strength
for the potential (equation \ref{eqn:vf})
\begin{equation}
\epsilon = { 2 \pi G S_m \Sigma_0 r \over |\alpha| }. \label{eqn:eps}
\end{equation}

The Hamiltonian  (equation \ref{eqn:KK}) is similar to
 that of a pendulum.  Shifting the angle $\theta'$ by a constant we can rewrite the Hamiltonian as 
 \begin{equation}
 K(J,\theta') =    {v_c^2 \over L_s^2} {J^2 \over 2} - \epsilon \cos(m\theta') 
 \end{equation}
 where we have neglected the constant term in the Hamiltonian, 
 assumed a flat rotation curve and flipped the sign
 from equation (\ref{eqn:KK}).
 The resonance width can be estimated by computing the energy 
 of the separatrix ($K=\epsilon$ at $J= 0$, $m\theta'=\pi$) and then
solving for $J$ at $\theta'=0$.
 The width of the resonance (peak to peak in $J$) is
\begin{equation}
\Delta J \approx 4 \sqrt{|\epsilon| \over v_c^2} L_s. \label{eqn:width}
\end{equation}
Here $\Delta J$ describes the width of the corotation resonance and determines
 the maximum radial distance that a particle can migrate in the resonance;
\begin{equation}
\Delta r \sim {\Delta J \over v_c} \sim 4 \sqrt{|\epsilon| \over v_c^2} r_s \label{eqn:deltar}
\end{equation}
where $r_s = L_s/v_c$.
Using equations (\ref{eqn:eps}) and (\ref{eqn:deltar}),
\begin{eqnarray}
\Delta r &\sim& 0.9 {\rm kpc}~  \left({ S_m \over 0.15 }\right)^{1\over 2} \left({r_s \over 8 {\rm kpc}}\right)^{3\over 2} 
\left( { \Sigma_0 \over 50 M_\odot {\rm pc}^2}\right)^{1 \over 2} \nonumber \\
&&
\left( { \alpha \over 17 }\right)^{-{1\over 2}}  \left({v_c \over 200 {\rm km~s}^{-1} } \right)^{-1} .
\end{eqnarray}
Here we have used a surface density similar to that of the Milky Way disk at the Sun's galactocentric radius \citep{holmberg04},
%The parameter $\alpha = 5.5$ for $m=2$ and a winding angle of $20^\circ$.
a parameter $\alpha = 17$ for $m=4$ and a winding angle of $13^\circ$ as preferred by some models for
the Milky Way \citep{vallee08}, and the 15\% density contrast estimated by \citet{drimmel01}.
This suggests that unless the Galaxy experienced strong and open spiral arms, migration over large
distances is unlikely.

Using the cosine perturbation we can also estimate the maximum migration rate using equations (\ref{eqn:Lmax},\ref{eqn:rmax}) and the derivative with respect to $\theta'$ of equation (\ref{eqn:vcos}),
\begin{eqnarray}
\dot r_{max} &\sim& {2 \pi G S_m \Sigma_0 r_s \over \beta} \nonumber \\
& \sim & 2 {\rm km~s}^{-1} \left({ S_m \over 0.15 }\right) \left({r_s \over 8 {\rm kpc}}\right)
\left( { \Sigma_0 \over 50 M_\odot {\rm pc}^2}\right) \nonumber \\
&& \left( { \beta \over 4 }\right)^{-1}  \left({v_c \over 200 {\rm km~s}^{-1} } \right)^{-1}, 
\end{eqnarray}
(here $\alpha = m \beta$).
Only if there are strong and open spiral arms is migration rapid.  
We see migration at a rate of about 20 km~s$^{-1}$ (see Figure \ref{fig:sig_rdot}) exceeding the
value above.  \citet{grand12b} show a migration rate (using their Figure 10) 
of about 1kpc in 20 Myr equivalent to 50 km~s$^{-1}$.
The
migration seen by \citet{grand12,grand12b} and us is both more extreme and quicker than
estimated above using parameters estimated for local Galactic spiral structure.
We will discuss this issue in more detail below.

For a particle to migrate the distance given in equation (\ref{eqn:deltar}), it must be captured as the pattern grows 
near $\phi \sim -\pi/2$ and at low $L$
and then released at $\phi \sim \pi/2$ at higher $L$, as described  by \citet{sellwood02}.
The dependence of the resonant angle on $\beta \ln L$ does not affect the resonance width estimate
(one can follow the same procedure to estimate the width of the separatrix).
However the shape and normalization of the function $f(\phi)$ does affect the resonance width.
%For a function similar to a cosine we can estimate
%\begin{equation}
%\Delta r \sim 4 \sqrt{|\epsilon f(\phi_c)| \over v_c^2} r_s \label{eqn:deltarf}
%\end{equation}  
%where $\phi=\phi_c$  in the stable fixed point of the resonance.  

How low an epicycle (eccentricity) is required for capture into a corotation resonance?  The eccentricity must
 be low enough that the angle, $\phi$, remains within $\pi/2$ of the potential gradient extremum, otherwise
 the torque on the particle would oscillate instead of remaining positive or negative, as exhibited by the
 migrating and non-migrating particles shown in Figure \ref{fig:rates}.
 We consider $\phi$ as a function of $r$ rather than angular momentum or
 $\phi = \beta \ln r - \theta'$.   An orbit with eccentricity $e$ has an apocenter of $r \sim r_0(1+e)$.
 Inserting the apocenter radius into the expression to $\phi$ we find $\phi \approx \beta \ln r_0 + \beta e$,
 consequently epicyclic variations cause a change in $\phi$ of order $\beta e$. 
 To keep $\phi$ from varying by more than $\pi/2$ as the particle movies radially in the epicycle we require that
\begin{equation}
 e \lesssim {\pi \over 2 \beta } = {\pi m\over 2 \alpha}. \label{eqn:elimit}
\end{equation}
The corotation resonance is likely to be ineffective for particle eccentricities above this value.
Tightly wound patterns can only cause migration in  extremely low eccentricity stars.

Previous studies have found that corotation resonances are only effective at causing migration
among low eccentricity particles (see Figure 12 by \citealt{minchev12} and associated discussion).
Using the $\alpha_i=6$ parameter for the $m=3$ pattern 
 from Table \ref{tab:tab1} listing our model patterns, equation (\ref{eqn:elimit}) gives a maximum 
eccentricity of 0.8 for particles to be captured into a corotation resonance.  
However in section \ref{sec:finding}, we estimated that only
particles with eccentricity $e \lesssim 0.2$ migrated, hence equation (\ref{eqn:elimit}) overestimates
the maximum particle eccentricity.    We will discuss this issue again when we consider
the Gaussian bar potential perturbation model below.

\subsection{Migration by a thin Gaussian bar}

When the potential perturbation is well approximated with a single Fourier component 
(equation \ref{eqn:vcos}, \ref{eqn:sigmacos}), then the maximum migration rate
occurs at an angle $\phi=\pi/2$.    However the density as a function of angle is poorly approximated
by single Fourier component (see Figure \ref{fig:peaks}) and we find that migration is taking place
when the particle is 30$^\circ$ (not $90^\circ$) of the density peak.  
Similarly the extreme migrators shown by \citet{grand12b} 
were within 10 -- 20$^\circ$ of a density peak (see their Figure 14).
We could expand the gravitational potential perturbation function  in terms of Fourier coefficients
with $f(\phi)  = \sum A_m \cos m\phi$.  However the maximum tangential force (and the angle 
at which the maximum is located) is then a function of the Fourier coefficients as the derivative of
each component reaches a maximum at a different angle.   

To take into account the narrowness of the spiral features, we  
 describe a local density peak as a bar-like linear feature
with a surface density profile (rather than in terms of Fourier components)
\begin{equation}
\Sigma (y) = \Sigma_p e^{-y^2/(2\sigma^2)},
\end{equation}
where $\Sigma_p$ is the peak density subtracted by the mean density, and the density bar has full-width-half-max of 2.35 $\sigma$.
This description has the advantage that it is a function of only three parameters,
the peak density, $\Sigma_p$, (above a background level), a width, described with $\sigma$, 
and an orientation angle.
We assume that the density feature does not strongly depend on the direction perpendicular to $y$.
We orient the density feature with the winding angle $\gamma$, with $\gamma=0$ corresponding to the feature
azimuthally oriented and $\gamma=90^\circ$ with the feature oriented radially.  Here $y$ increases
in the direction perpendicular to the feature.
For a tightly wound structure $\beta^{-1} = \tan \gamma \sim \gamma$.

The Fourier transform of the above density profile is also Gaussian with 
$\Sigma(k) = \Sigma_p \sigma \exp(-k^2 \sigma^2/2)$.
In the thin disk approximation, and using Poisson's equation, each Fourier coefficient of the 
gravitational potential $\propto e^{iky - |kz|}$.  Using a pillbox about the plane we can show that 
for each Fourier coefficient the gravitational potential\footnote{The integral 
of $\Phi(k) dk$ does not converge, however the integral should be cut off at 
small $k$ due to the finite thickness of the disk. 
% It is something like
% $\Phi(y) = \int_0^\infty \sqrt{2 \pi} G\Sigma_p \exp -(k^2 \sigma^2/2) \cos(ky) k^{-1} dk$.  
% The problem is at small $k$ which corresponds to large size
%scales so this is not really a problem.  You need to cut it off at small k.
}
\begin{equation}
\Phi(k) = - {\sqrt{2 \pi} G\Sigma_p \sigma \over |k|} \exp \left({-k^2\sigma^2\over 2} + iky \right).
\end{equation} 
Taking the $y$ derivative and integrating over $k$
\begin{eqnarray}
{\partial \Phi \over \partial y} &=& 2 \sqrt{2\pi} G \Sigma_p \int_0^\infty e^{-k^2/2} \sin \left({k y \over \sigma }\right) dk\\
&=& 4 \sqrt{\pi} G \Sigma_p ~{\rm DawsonF} \left({x \over \sqrt{2} \sigma} \right),
\end{eqnarray}
where the Dawson-F integral is defined as 
\begin{equation} F(x) \equiv \exp( -x^2) \int_0^x \exp y^2 dy.\end{equation}
The Dawson integral peaks approximately at $F(1) \sim 0.5$, consequently
the maximum drift rate should occur for a particle located at a distance $x  \sim \sqrt{2} \sigma$ 
from the peak or approximately at at the half-width-half max position.

We can now consider a rotating logarithmic Gaussian bar with surface density
\begin{equation}
\Sigma (r, \theta, t) = \Sigma_p \exp \left( - {( \beta \ln r -  (\theta- \Omega t))^2 \over 2 s^2} \right)
\end{equation}
with angular width $s$. We can relate $s=\sigma/r$.
For a Gaussian bar tilted at an angle $\gamma$ we can approximate locally
\begin{equation}
{\partial \Phi \over \partial \theta }  \sim {\partial \Phi \over \partial y} \sin \gamma.
\end{equation}
The angular derivative of the gravitational potential  
\begin{equation}
{\partial \Phi \over \partial \theta }  \sim  4 \sqrt{\pi} G \Sigma_p \sin \gamma ~{\rm DawsonF} \left({\phi \over \sqrt{2} s} \right)
\end{equation}
with $\gamma$ the winding angle and $\phi $ defined as in equation (\ref{eqn:phi}).

Using the derivative $d\Phi/d \theta$, we can estimate the maximum torque $\tau_{max}$ and from this the maximum
drift rate $\dot r \sim \tau_{max}/v_c$ or 
\begin{equation}
{\dot r}_{max} \sim G \Sigma_p 2 \sqrt{\pi} \Omega^{-1} \sin \gamma
\end{equation}
or
\begin{eqnarray}
%{dr \over dt} v_c^{-1} & \sim & 0.06 
{\dot r}_{max}  & \sim & 12 {\rm km~s}^{-1} 
\left({\Sigma_p \over 50 M_\odot {\rm pc}^{-2}} \right) \left( {\sin \gamma  \over 24^\circ} \right) \times\nonumber \\ &&
\left( { r_0 \over 8 {\rm kpc}}\right) \left( { v_c \over 200 {\rm km~s}^{-1}}\right)^{-1}. \label{eqn:rate}
\end{eqnarray}
Here we have used the typical $24^\circ$ winding angle estimated for the spiral features in the simulation \citep{quillen11}
and a surface density similar to that of the Milky Way disk at the Sun's galactocentric radius \citep{holmberg04}.
The above estimate suggests that high density peaks are most effective at causing migration.
The migration rate is only fast enough to cause significant migration within an orbital period if the
surface density is of order $50 M_\odot$ pc$^{-2}$.  As mentioned previously, this presents a potential problem for
migration models for the Milky Way as peak spiral amplitudes have been estimated
at levels much below this (e.g., \citealt{drimmel01}).

The maximum drift rate occurs for a particle
approximately at the half-width-half-max location and so is sensitive to
the peak width.   The particle must remain within a half-width of this location, hence the upper limit
on eccentricity for migration particles must be modified from that estimated using
a cosine potential perturbation (equation \ref{eqn:elimit}).
Using an apocenter radius $r = r_0 (1 + e)$ and keeping $\phi$ within angular width $s$ of the maximum migration
rate position we estimate that for a particle to migrate 
\begin{equation}
e \lesssim {s \over \beta}.
\end{equation}
For $\beta \sim 2$ in our model (taking values from Table \ref{tab:tab1}) and width $s \sim 30^\circ \approx 0.5$
radians
from Figure \ref{fig:peaks} we estimate $e\lesssim 0.25$ and this is approximately consistent with our previous 
estimate (based on Figure \ref{fig:movie}a) that the maximum migrator eccentricity $e_{max} \sim 0.2$.

\begin{figure}
\includegraphics[width=3.6in, trim=0.9in 0.2in 0.5in 0in]{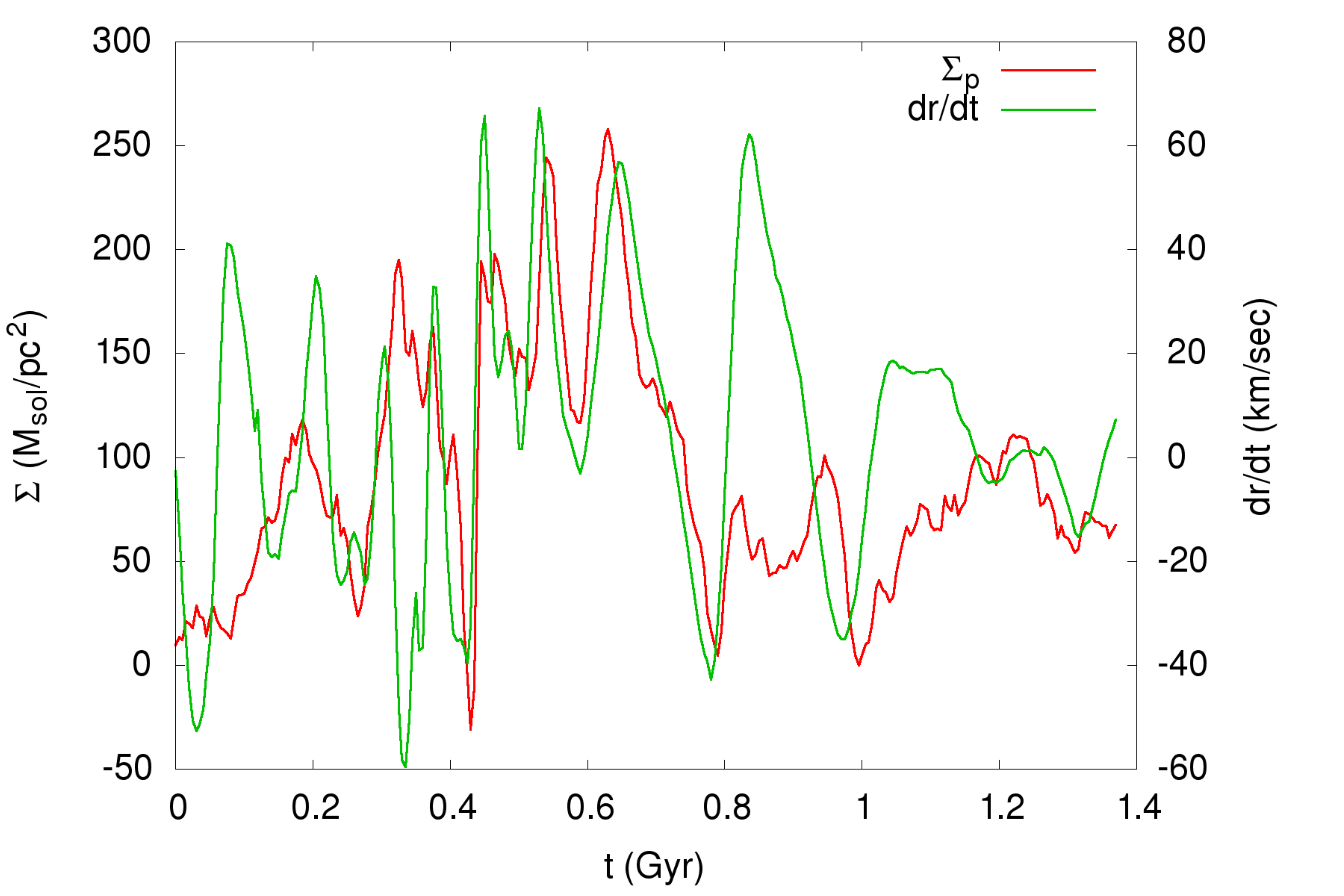}
\caption{
Peak density and migration rate.
For the selected particle shown in 
Figure \ref{fig:rates}a
the nearest peak mass surface density density, $\Sigma_p$, subtracted by the mean density, is shown 
as a red line and has  $y$-axis on the left.   Also shown as a green line
and with  along with $y$-axis on the right, is the rate of radial migration,
$\dot r$.  Both $\Sigma_p$ and $\dot r$ are shown as a function of time.
During the period of migration from 500 to 700 Myr,
the peak density and migration rate are correlated, as we would expect from
a corotation resonance mediated migration model. 
The rate of migration and peak density are approximately consistent with the maximum
estimated using the Gaussian bar model (equation \ref{eqn:rate}).
}
\label{fig:sig_rdot}
\end{figure}

Our estimate for the migration rate (equation \ref{eqn:rate}) depends on the strength of the potential term 
and this is set by the peak surface density of the feature.
Using the Gaussian bar model we compare  estimated migration rates to those
measured in the simulation.  In Figure \ref{fig:sig_rdot}
we show both the migration rate and density of the nearest spiral peak for our
selected particle (with trajectory also illustrated in Figure \ref{fig:movie}a, \ref{fig:peaks}, and \ref{fig:rates}). 
As expected, the migration rate is tightly correlated with the peak density of the nearest spiral
feature.
During the period of migration from 500 Myr to 700 Myr, the density $\Sigma_p$ of the nearest density peak
has values in the range of 100--$250 M_\odot {\rm pc}^{-2}$. 
Using a mean value of $\Sigma_{p} \approx 180 M_\odot $ pc$^{-2}$ and equation (\ref{eqn:rate}) 
we estimate a migration rate of 40 km~s$^{-1}$.  This is
somewhat higher than the mean migration of about 20--30 km~s$^{-1}$, but we should remember that
equation (\ref{eqn:rate}) estimates a maximum value and so we expect the actual value 
to be somewhat lower. 

%Figure \ref{fig:peaks} suggests that the width of the spiral features ranges from 20--40$^\circ$.
%This Figure also implies that rapid migration primarily takes place with the particle located
%approximately a half width away from the peak, as expected from the Gaussian bar model.

We have found that the Gaussian bar model successfully estimates the migration rate,
 the limiting eccentricity value for migration and the angular position of a particle that is 
 rapidly migrating.  In these ways the Gaussian bar model is superior to a model
 that approximates the potential perturbation with a a single Fourier component. 
 However, even though the derivative of the potential
perturbation is well defined for the Gaussian bar, the potential itself is not, making it impossible to modify 
equation
(\ref{eqn:deltar}) to estimate the extent that a particle can migrate.
Up to this point we have not considered how growth and variations in winding angle
effect the migration model.  However if migration is rapid, then a local approximation for
the Gaussian bar is sufficient to estimate the migration.   To estimate the extent that
a particle can migrate we would have to take into account the time dependent form 
of the potential perturbation and the particle's trajectory (in $\phi$) with time.

\section{Discussion and Summary}

As have other studies \citep{grand12,grand12b,minchev12}, we find stars that migrate in radius
in our N-body simulation.  However spectrograms of the simulation do not show transient patterns that are strong
near their corotation radius, presenting a difficulty for the corotation resonance model 
for migration \citep{sellwood02}.   
 We find that particles migrate outwards  in the simulation
when they are at low eccentricity and lag a short lived local density peak, confirming the results of previous studies 
\citep{grand12,grand12b}.  
Density peaks may not survive more
than a rotation period,  suggesting that they are not due to individual growing and decaying spiral density waves. 
We consider the possibility that they are due to interference between density waves. 
This scenario does not
require that spiral density waves last with fixed amplitudes, pattern speeds or winding angles 
for dozens of rotation periods and so is not inconsistent with recent studies
that suggest that spiral structure is transient (e.g., \citealt{sellwood11}).

We have explored a simple model for interfering spiral density waves and find that
they can produce short lived local density peaks and these, if they are sufficiently strong, can account for
the migration seen in our simulation.  The mechanism for migration due to 
a corotation resonance proposed by \citet{sellwood02} can be modified to apply for a short-lived local density enhancement.  Using a Gaussian bar model for the potential perturbation, estimates
of the migration rate, angular offset between particle and spiral feature, and maximum eccentricity
for migrators roughly agrees with the values measured in the simulation.
%We estimate the rate of migration that can be caused by  such a peak and 
%find that it is approximately proportional to the difference
%between the feature's peak and mean surface density.     
We show that interference peaks can appear to rotate
faster at smaller radii than outer radii (and so appear to wind up) as seen in our and other simulations
\citep{grand12}.   The angular rotation rate of an interference peak should lie between the pattern speeds
of the two patterns that cause it.    Nearly corotating interference peaks would be particularly likely
to appear near the end of a bar  due to interference between the fast bar and more slowly moving spiral patterns
outside the bar.  This may account for ongoing radial migration that is seen in simulations after bar formation 
and in the vicinity of the bar.

We can contrast a migration mechanism mediated by individual spiral density waves (as proposed
by \citealt{sellwood02}) and one mediated
by short lived interference peaks.    An interference peak can have a higher surface density than each individual pattern.  
As the distance migrated and speed of migration depends on the surface density, interference peaks can enhance migration.
Many interference peaks can be produced over the lifetime of a few spiral density waves.  This suggests that
migration events could occur faster and more frequently than when considering a model where migration events
only occur when a single pattern grows and decays.   This may ameliorate the timescale problem as only a few dozen spiral
density waves could arise and decay during the age of the Galactic disk, however a few times this number of interference
peaks may arise during the same time if there are multiple spiral density waves present.
With a single pattern, migration can only occur at the corotation resonance of the pattern.   Interference peaks can
have a variety of effective angular rotation rates, depending upon the amplitudes and patterns of the waves causing them.
Consequently migration induced by  interference peaks may be more pervasive than migration induced by 
individual spiral density waves.

The migration rate and the distance a star can migrate
 due to corotation resonance are dependent on the peak density in a spiral wave.  As the stellar surface density 
decreases with increasing radius, migration due to corotation resonances may be less effective in the outer parts
of the Galaxy.   Furthermore as migration likely occurred near the Sun's radius it is likely that spiral structure
is either stronger than previously estimated (e.g., by \citealt{drimmel01}), 
or was stronger in the past than currently observed in the Galaxy.
Better measurements of the spiral structure in the Galaxy are needed to make quantitative estimates
of the extent of on-going radial migration.

% Short lived peaks are more likely to be due to interference between patterns rather than rapid growth and decay of patterns. 
 Interference between spiral 
density waves can mimic  transient-like behavior, including inducing migration as discussed here.
Short lived density peaks can also cause localized bursts of star formation that can move across a galaxy.
This picture of star formation differs from that expected with a single strong density wave that would produce a constant
amplitude wave of star formation.   We estimate that the rate that a burst of star formation moves across a galaxy depends on the
winding angle of the two patterns.  A burst would move inwards if the outer galaxy contains more tightly wound
spiral structure than the inner galaxy.  Future studies could probe for evidence of localized bursts
of star formation in context with spiral structure models to better constrain
the nature of spiral structure in galaxies.  Future studies can also aim to quanitify the statistics of migrating stars
using interference mediated corotation migration models.

\vskip 0.3 truein
Acknowledgements.   
This work was in part supported by NSF through award AST-0907841.   
We thank NVIDIA for gifts of graphics cards.  We thank Jerry Sellwood for helpful comments.

\end{document}